\documentclass[prb,
                    preprint
                    ]{revtex4}

\usepackage{epsfig}

\def\e{\begin{equation}}
\def\f{\end{equation}}
\def\ea{\begin{eqnarray}}
\def\fa{\end{eqnarray}}
\def\=#1{\overline{\overline{#1}}}
\def\_#1{{\bf #1}}

\def\.{\cdot}

\def\l#1{\label{eq:#1}}
\def\r#1{(\ref{eq:#1})}

\def\l#1{\label{eq:#1}}
\def\r#1{(\ref{eq:#1})}

\begin{document}

\title{Experimental verification of the key properties
of a three-dimensional isotropic transmission line based superlens}

\author{Pekka Alitalo}

\author{Stanislav Maslovski}

\author{Sergei Tretyakov}

\affiliation{Radio Laboratory / SMARAD, Helsinki University of Technology\\
P.O. Box 3000, FI-02015 TKK, Finland\\
{\rm E-mails: pekka.alitalo@tkk.fi, stanislav.maslovski@gmail.com,
sergei.tretyakov@tkk.fi}}

\date{\today}

\begin{abstract}

Design and experimental realization of a three-dimensional
superlens based on {\it LC}-loaded transmission lines are
presented. Commercially available components and materials are
used in the design. Transmission properties of the designed
structure are studied experimentally and the observed lens
properties are compared with analytical predictions. Backward-wave
propagation and amplification of evanescent waves in the prototype
structure are verified both analytically and experimentally.

\end{abstract}

\maketitle

\section{Introduction}

Systems that are able to focus propagating electromagnetic waves
and amplify evanescent waves have received a lot of attention
after Pendry published his paper\cite{Pendry} about a superlens
capable of subwavelength focusing. The amplification of evanescent
waves is the key feature in the subwavelength focusing
characteristics of the superlens, because evanescent waves carry
information about fine details (smaller than the wavelength) of
the source field. Pendry's superlens is based on a planar slab
made of a backward-wave (BW) material (also called double-negative
material or Veselago medium), in which the real parts of the
effective permittivity and permeability are both negative.

The first successful demonstrations of a BW material and negative
refraction were done using an array of resonant cells, which were
comprised of thin wire strips (effective negative permittivity)
and split-ring-resonators (effective negative permeability).
\cite{Smith,Shelby,Parazzoli} However, these structures are highly
anisotropic and they allow one to achieve the properties of a BW
material only within a narrow frequency band due to the use of
resonant phenomena in split rings. Also, such systems would be
very difficult to implement at RF-frequencies. This is why other
ways to realize a BW material have been extensively studied in the
recent literature. Another approach to create a BW material is
based on {\it LC}-loaded transmission-line
networks.\cite{Eleftheriades,Caloz} These networks do not rely on
resonant response from particular inclusions, and the periods of
the structures can be made very small as compared to the
wavelength. These features allow realization of broadband and
low-loss devices, which is extremely difficult if resonant
inclusions are used.

So far the transmission-line (TL) network approach has been
successfully realized in one- and two-dimensional
networks,\cite{Caloz2,Grbic3} and the main challenge on the route
towards truly three-dimensional broadband and low-loss superlenses
is the realization of isotropic three-dimensional artificial BW
structures.  In our recent paper we introduced a three-dimensional
(3D) {\it LC}-loaded transmission-line network and derived
necessary equations to design such a structure.\cite{Alitalo} The
key idea of our approach is that the inside volume of every TL
section is effectively screened from the other sections and from
the outer space.\cite{Alitalo} Other ways to design
three-dimensional FW and BW transmission-line structures have been
proposed,\cite{Grbic4,Hoefer} but to the best of our knowledge
those structures have not been realized.

The goal of this paper is to show that the three-dimensional
structure analytically described earlier\cite{Alitalo} can be
practically manufactured and allows us to realize the two basic
properties of Pendry's superlens, i.e. backward-wave propagation
and amplification of evanescent waves.

\section{The designed structure}

The components that are used to realize the 3D superlens are
listed in Table~\ref{table1} with their main properties. The
structure is designed to work at a frequency close to 1 GHz. The
period of the structure ($d$) should be much smaller than the
wavelength ($\lambda$) at this frequency. Accordingly, the period
of the structure is chosen to be 13 mm ($\lambda_0$ at 1 GHz is
300 mm, $\lambda_0$ is the wavelength in free space). For the ease
of manufacturing, the thickness of the BW slab ($l$) has been
decided to be three periods, i.e. $l=3d=39$ mm. Using the
equations for the characteristic impedances,\cite{Alitalo}
suitable impedance values of the TLs have been found to be $Z_{\rm
0,TL,FW}=66$ $\Omega$ (impedance of the TLs in the FW region) and
$Z_{\rm 0,TL,BW}=89$ $\Omega$ (impedance of the TLs in the BW
region). With these values the characteristic impedances of the FW
and BW regions should be approximately equal. The height of the
substrate is 0.787 mm (Rogers RT/Duroid 5870). The lumped
elements, i.e., the capacitors and inductors, have been supplied
by American Technical Ceramics Corp. (ATC).

\begin{table}[h]
\centering \caption{Properties of the prototype components.}
\label{table1}
\begin{tabular}{|c|c|c|c|c|} \hline
   &  Value   &  Tolerance  &  Manufacturer  & Q-factor or\\[-0.7em]
   &          &             &                & loss tangent\\
\hline

$d$ & $13$ mm & - & - & - \\

$Z_{\rm 0,TL,FW}$ & $66$ $\Omega$ & - & - & -\\

$Z_{\rm 0,TL,BW}$ & $89$ $\Omega$ & - & - & -\\

$C$  &  $3.3$ pF &  $\pm0.05$ pF  & ATC  & $Q_{C,\rm 1 GHz}=500$\\

$L$  &  $6.8$ nH &  $\pm0.136$ nH  & ATC  & $Q_{L,\rm 1 GHz}=50$\\

Substrate  &  $\varepsilon_r=2.33$ &  $\pm0.02$ &  Rogers & $\tan\delta=0.0012$\\
\hline
\end{tabular}
\end{table}

Using the well-known equations for microstrip lines, one can find
the widths ($w$) and the effective permittivities
($\varepsilon_{r,{\rm eff}}$) for the required transmission lines
(66 $\Omega$ and 89 $\Omega$). The results for the FW network are
$w_{\rm FW}\approx 1.437$ mm, $\varepsilon_{r,{\rm eff},\rm
FW}\approx 1.902$, and for the BW network $w_{\rm BW}\approx
0.794$ mm, $\varepsilon_{r,\rm eff,BW}\approx 1.845$.

\begin{figure}[h]
\centering \epsfig{file=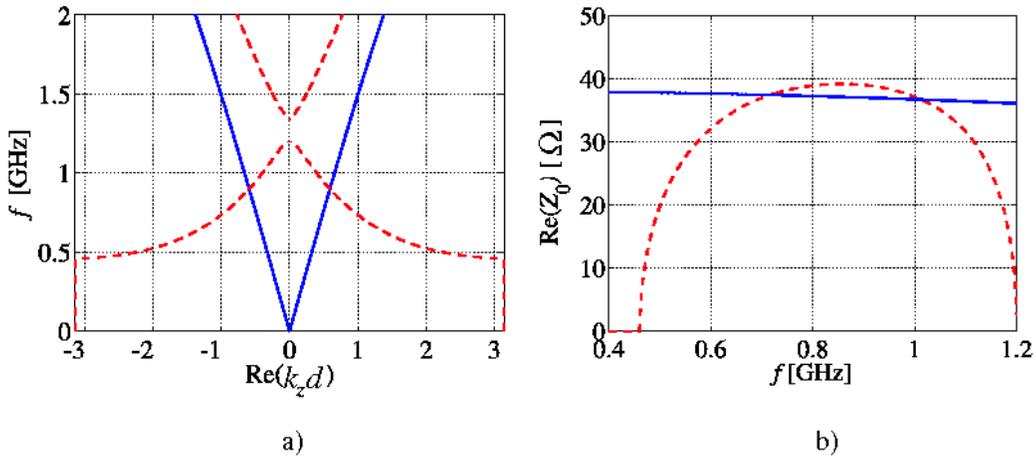, width=0.9\textwidth}
\caption{a) Dispersion curves for the FW and BW networks (ideal,
lossless components). b) Characteristic impedances for the FW and
BW networks (ideal, lossless components). Propagation along the
$z$-axis is considered. Solid lines: FW network, dashed lines: BW
network.} \label{dispersion_Z0_fvar}
\end{figure}

Detailed description of the dispersion equations and
characteristic impedances for the FW and BW networks will be
published elsewhere\cite{Alitalo} and due to their complexity they
are not repeated here. By plotting the dispersion curves for the
FW and BW networks, one can see that the matching frequency (at
which the longitudinal component of the wavenumber in the BW
region equals to the negative of the longitudinal component of the
wavenumber in the FW region, i.e. $k_{z,\rm BW}=-k_{z,\rm FW}$) is
$f\approx 0.8996$ GHz, see Fig.~\ref{dispersion_Z0_fvar}a. In the
dispersion curves plotted here only propagation along the $z$-axis
(lens axis) is considered. From Fig.~\ref{dispersion_Z0_fvar}a it
is also seen that at the matching frequency the absolute value of
$k_z d$ is approximately 0.5909 (in both FW and BW regions), which
corresponds to the longitudinal wavenumber $k_z=0.5909/d\approx
45.5$ m$^{-1}$. This is equal to the maximum transverse wavenumber
($k_{\rm t}$) that a propagating wave can have, and therefore we
can conclude that for evanescent waves $k_{\rm t}>45.5$ m$^{-1}$.
The maximum transverse wavenumber for evanescent waves is at the
edge of the first Brillouin zone:
 $k_{\rm t,max}=\pi/d\approx 242$ m$^{-1}$.
By plotting the characteristic impedances ($Z_0$) for the FW and
BW networks, one can see that the characteristic impedances of the
two networks are approximately equal at the design frequency
($f\approx 0.8996$ GHz), see Fig.~\ref{dispersion_Z0_fvar}b.

\section{Transmission properties of the designed structure}

\begin{figure}[h]
\centering \epsfig{file=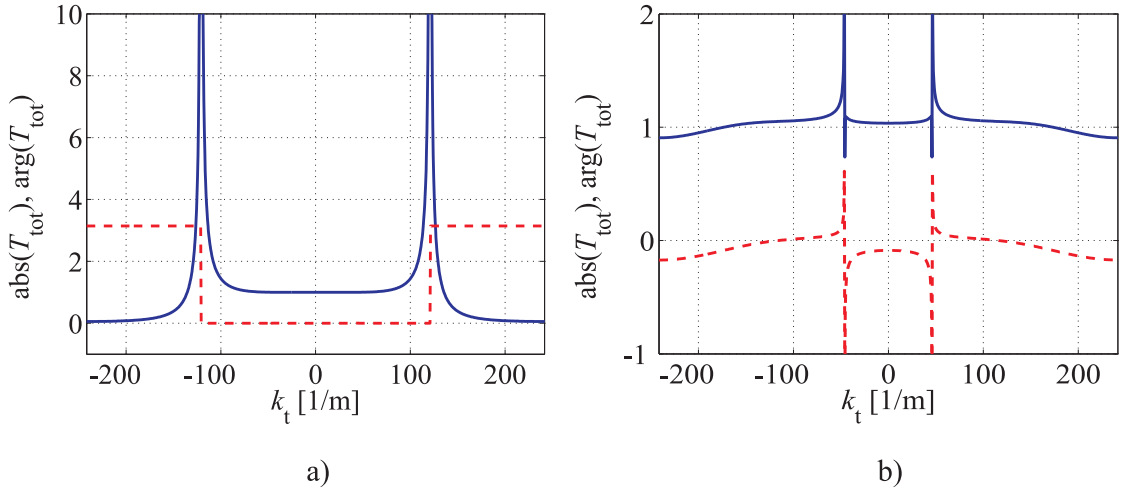, width=0.9\textwidth}
\caption{Absolute value (solid line) and phase (dashed line) of
the transmission coefficient of the designed superlens as a
function of the transverse wavenumber. a) $f=0.8996$ GHz, ideal
components. b) $f=0.91$ GHz, dissipation in the lumped components
and in the substrate taken into account.}\label{Ttot_ktvar1and4}
\end{figure}

The equation for the transmission coefficient of the lens ($T_{\rm
Lens}$) as a function of the transverse wavenumber $k_{\rm t}$ was
derived earlier\cite{Alitalo}, and the result was: \e T_{\rm
Lens}(k_{\rm t})=\frac{4Z_{\rm 0,FW}(k_{\rm t})Z_{\rm 0,BW}(k_{\rm
t})}{[Z_{\rm 0,FW}(k_{\rm t})+Z_{\rm 0,BW}(k_{\rm
t})]^{2}e^{+jk_{z,\rm BW}(k_{\rm t})l}-[Z_{\rm 0,FW}(k_{\rm
t})-Z_{\rm 0,BW}(k_{\rm t})]^{2}e^{-jk_{z,\rm BW}(k_{\rm t})l}},
\l{T}\f where we again assume that the lens axis is parallel to
the $z$-axis. The total transmission from the source plane to the
image plane is then (the distance from the source plane to the
lens is $s_1$, and the distance from the lens to the image plane
is $s_2$)\cite{Alitalo} \e T_{\rm tot}(k_{\rm t})=T_{\rm
Lens}(k_{\rm t})e^{-jk_{z,\rm FW}(k_{\rm t})\cdot(s_1+s_2)}.
\l{T_tot} \f To estimate the performance of the designed
superlens, the total transmission from the source plane to the
image plane can be plotted using equations \r{T} and \r{T_tot},
see Fig.~\ref{Ttot_ktvar1and4}a, where we have used $s_1=s_2=19.5$
mm and $l=39$ mm (lossless case).

Tuning the frequency, imaging can be improved, effectively
enhancing transmission of the modes with $k_{\rm t}>45.5$ m$^{-1}$
(i.e., evanescent waves). This is due to a better matching of the
characteristic impedances at those values of $k_{\rm t}$ that
correspond to evanescent waves. For example, in the lossless case
at $f=0.91$ GHz the transmission coefficient $T_{\rm tot}$ is
practically equal to unity for all transverse wavenumbers in the
range $0 \le k_{\rm t} \le k_{\rm t, max}$. The effect of
dissipation caused by the substrate and the lumped components can
be considered by taking into account the loss tangent ($\tan
\delta$) of the substrate and the quality factors ($Q$) of the
lumped components (shown in Table~\ref{table1}), see
Fig.~\ref{Ttot_ktvar1and4}b for this case. As is seen from
Fig.~\ref{Ttot_ktvar1and4}b, the transmission properties of this
lossy structure are close to the ideal case: $|T_{\rm tot}|\approx
1$ and $\arg(T_{\rm tot})\approx 0$ for a wide range of $k_{\rm
t}$ which includes propagating as well as evanescent spectral
components.

\section{Two-dimensional prototype} \label{sec_2D_prototype}

\begin{figure}[h]
\centering \epsfig{file=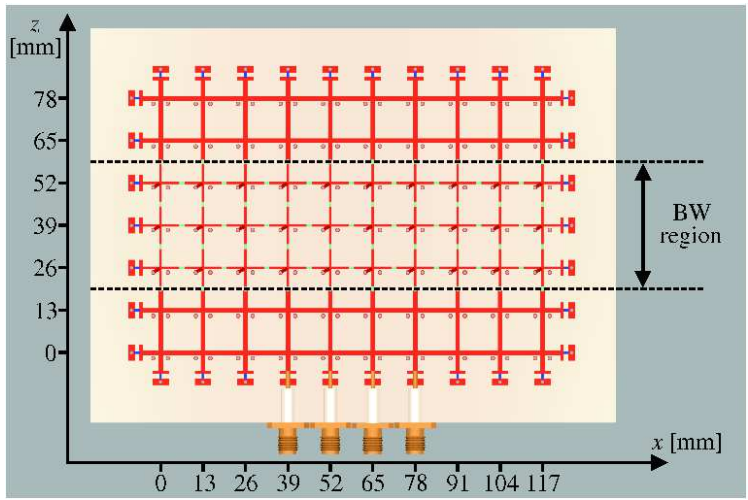, width=0.6\textwidth}
\caption{(Color online). Two-dimensional prototype of the proposed
structure (one horizontal layer of the designed 3D
structure).}\label{1_layer}
\end{figure}

First, in order to check the operational principles of the
proposed structure, a two-dimensional prototype was built (see
Fig.~\ref{1_layer}). The prototype had the same properties and
component values as shown in Table~\ref{table1}.  The edges of the
structure were terminated with resistive loads that were
approximately matched to the TL impedances. This was done in order
to reduce reflections of the propagating modes from the edges of
the structure.

The structure was excited by a coaxial feed (SMA-connectors)
connected to the edge of the first FW region as shown in the
bottom of Fig.~\ref{1_layer}. To have a possibility to change the
position of the excitation, four SMA-connectors were soldered to
the structure. The inner conductors of the SMA-connectors were
soldered to the microstrip lines and the outer conductors to the
ground. The SMA-connectors that were not used at each measurement
were terminated with 50 $\Omega$ loads.

\begin{figure}[h]
\centering \epsfig{file=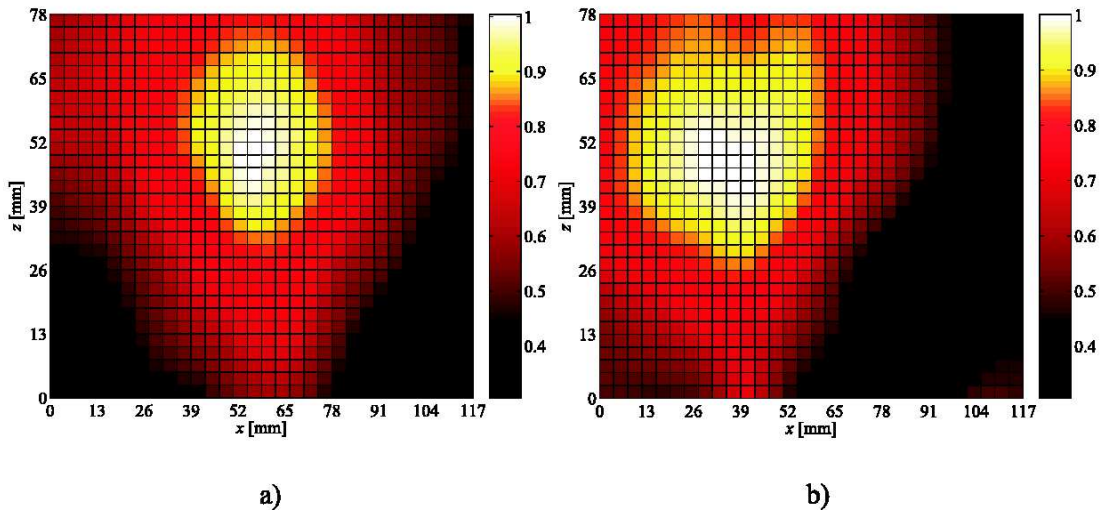, width=0.9\textwidth}
\caption{(Color online). Measured amplitude of the vertical
component of electric field on top of the 2D structure at $f=900$
MHz. Fields normalized to the maximum value. a) Symmetrical
excitation by two sources at $x=52$ mm, $z=-6.5$ mm and $x=65$ mm,
$z=-6.5$ mm. b) One source at $x=39$ mm, $z=-6.5$
mm.}\label{amplitude_1}
\end{figure}

By connecting port 2 of a vector network analyzer to the
excitation point(s) of the structure and port 1 to a probe antenna
(a short vertical monopole antenna), the electric field
distribution on top of the structure could be measured (by
measuring $\rm S_{12}$). The measured vertical component of the
electric field is proportional to the voltage at the network
nodes, a non-invasive direct measurement of which can be a
complicated task at 1 GHz. The probe antenna was connected to an
automated measurement robot, which could be programmed to position
the probe at certain points. Here the field was measured at the
center of each node of the structure which corresponds to 70
measurement points. The BW region is situated in the area $19.5$
mm $<z<58.5$ mm. See Fig.~\ref{amplitude_1} for the measured
electric field distributions on the top of the structure.

As is seen in Fig.~\ref{amplitude_1}a and Fig.~\ref{amplitude_1}b,
the maximum values of the amplitude occur at the back edge of the
BW region (as it is expected from the theory). In
Fig.~\ref{amplitude_1}b the point of excitation is displaced from
the middle to show that the effects seen are not caused by
reflections from the side edges. It is clear that both propagating
and evanescent modes are excited in the structure, because the
fields do not experience significant decay in the first FW region
(evanescent modes decay exponentially) and there is a remarkable
growth of the amplitude in the BW region (only evanescent modes
can be ``amplified'' in a passive structure like this). The
experiment does not show any noticeable reflections at the FW/BW
interfaces, which implies to a good impedance matching between the
two types of networks.

\begin{figure}[h]
\centering \epsfig{file=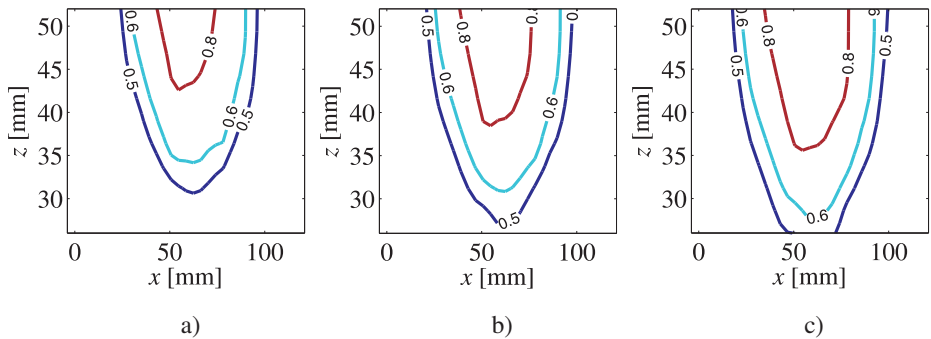, width=0.9\textwidth}
\caption{(Color online). Time-harmonic electric field on top of
the 2D structure at $f=900$ MHz. Fields normalized to the maximum
value. Symmetrical excitation by two sources at $x=52$ mm,
$z=-6.5$ mm and $x=65$ mm, $z=-6.5$ mm. a) $\phi=1$, b)
$\phi=1+\pi/20$, c) $\phi=1+2\pi/20$.}\label{contours_1_2}
\end{figure}

To show that the structure supports backward waves, the
time-harmonic electric field was plotted from the measured complex
field using \e E_{\rm real}={\rm Re}\{E_{\rm complex}e^{j\omega
t}\}. \f When the field plot is animated as a function of time, it
is seen that the waves propagate ``backwards'' (towards the point
of excitation) in the BW region. To illustrate this effect, some
contour lines of the time-harmonic field are plotted in
Fig.~\ref{contours_1_2} with different values of the phase angle
($\phi=\omega t$).

\section{Three-dimensional realization}

\begin{figure}[h]
\centering \epsfig{file=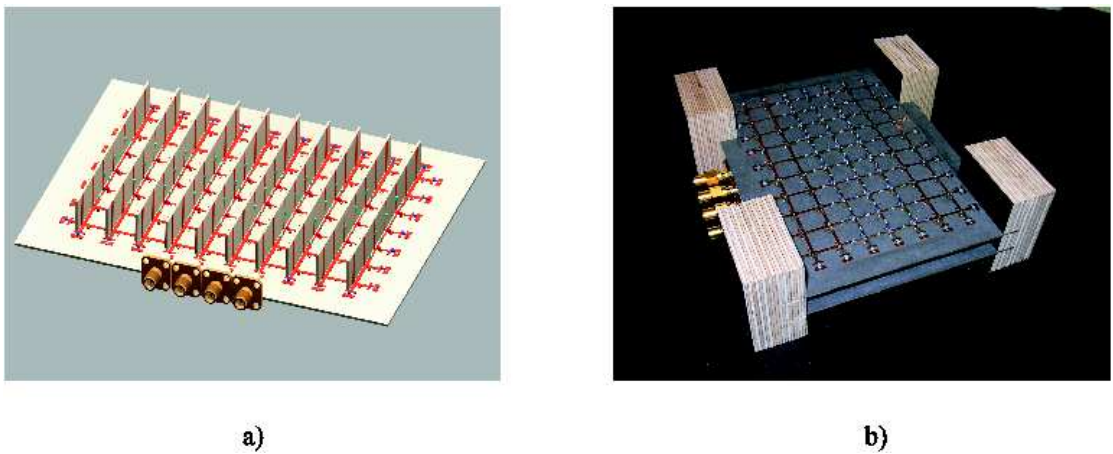, width=0.9\textwidth}
\caption{(Color online). a) Geometry of the 3D structure (one
horizontal layer and ten vertical sub-layers shown). b)
Experimental prototype of the 3D structure with two horizontal
layers.}\label{2_layer}
\end{figure}

To realize a three-dimensional structure, a second two-dimensional
layer as in section~\ref{sec_2D_prototype} was manufactured. To
connect these two layers, ten vertical sub-layers of height 12.2
mm were soldered between them. See Fig.~\ref{2_layer}a for the
geometry of the structure (one horizontal layer and ten vertical
sub-layers are shown). The resulting 3D structure is isotropic
with respect to waves propagating inside the TLs (distance between
adjacent horizontal and vertical nodes remains the same and the
vertical microstrip lines are also loaded with capacitors in the
BW region). A photograph of the manufactured two-layer structure
is shown in Fig.~\ref{2_layer}b.

\begin{figure}[h]
\centering \epsfig{file=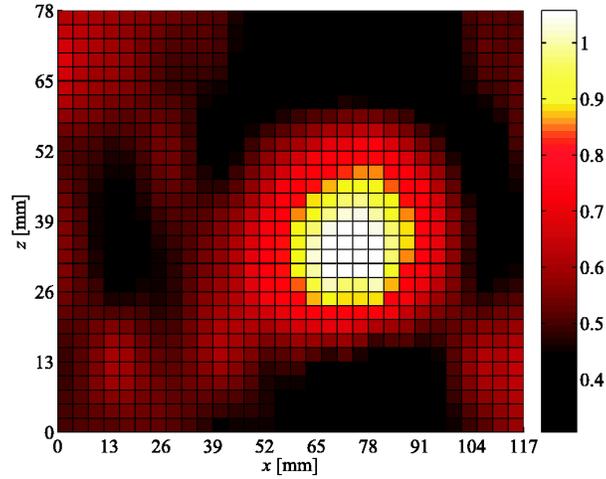, width=0.5\textwidth}
\caption{(Color online). Measured amplitude of the vertical
component of electric field on top of the 3D structure (two
horizontal layers) at $f=900$ MHz. Fields normalized to the
maximum value. One source below the lower horizontal layer at
$x=78$ mm, $z=39$ mm.}\label{amplitude_2_b}
\end{figure}

\begin{figure}[h]
\centering \epsfig{file=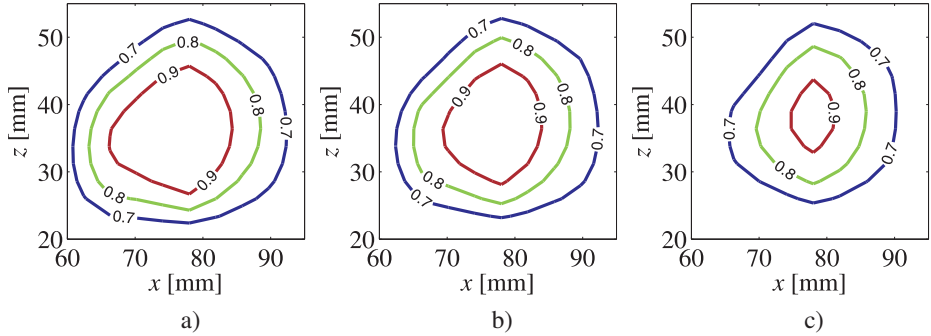, width=0.9\textwidth}
\caption{(Color online). Time-harmonic electric field on top of
the 3D structure (two horizontal layers) at $f=900$ MHz. Fields
normalized to the maximum value. One source below the lower
horizontal layer at $x=78$ mm, $z=39$ mm. a) $\phi=1.2$, b)
$\phi=1.2+\pi/20$, c) $\phi=1.2+2\pi/20$.}\label{contours_2_b}
\end{figure}

Having more than one layer in the structure, wave propagation
along the vertical axis (the $y$-axis) can be also experimentally
tested. This was done by exciting the structure from the bottom of
the lower horizontal layer at $x=78$ mm, $z=39$ mm. See
Fig.~\ref{amplitude_2_b} for the electric field distribution
measured on top of the upper layer and Fig.~\ref{contours_2_b} for
the instantaneous electric field snapshots.
Fig.~\ref{amplitude_2_b} proves the three-dimensional isotropy of
the proposed network that was predicted
theoretically.\cite{Alitalo} Fig.~\ref{contours_2_b} demonstrates
backward-wave propagation, because the point of the source appears
to be a ``sink'' for moving contours of instantaneous values of
the electric field.

\begin{figure}[h]
\centering \epsfig{file=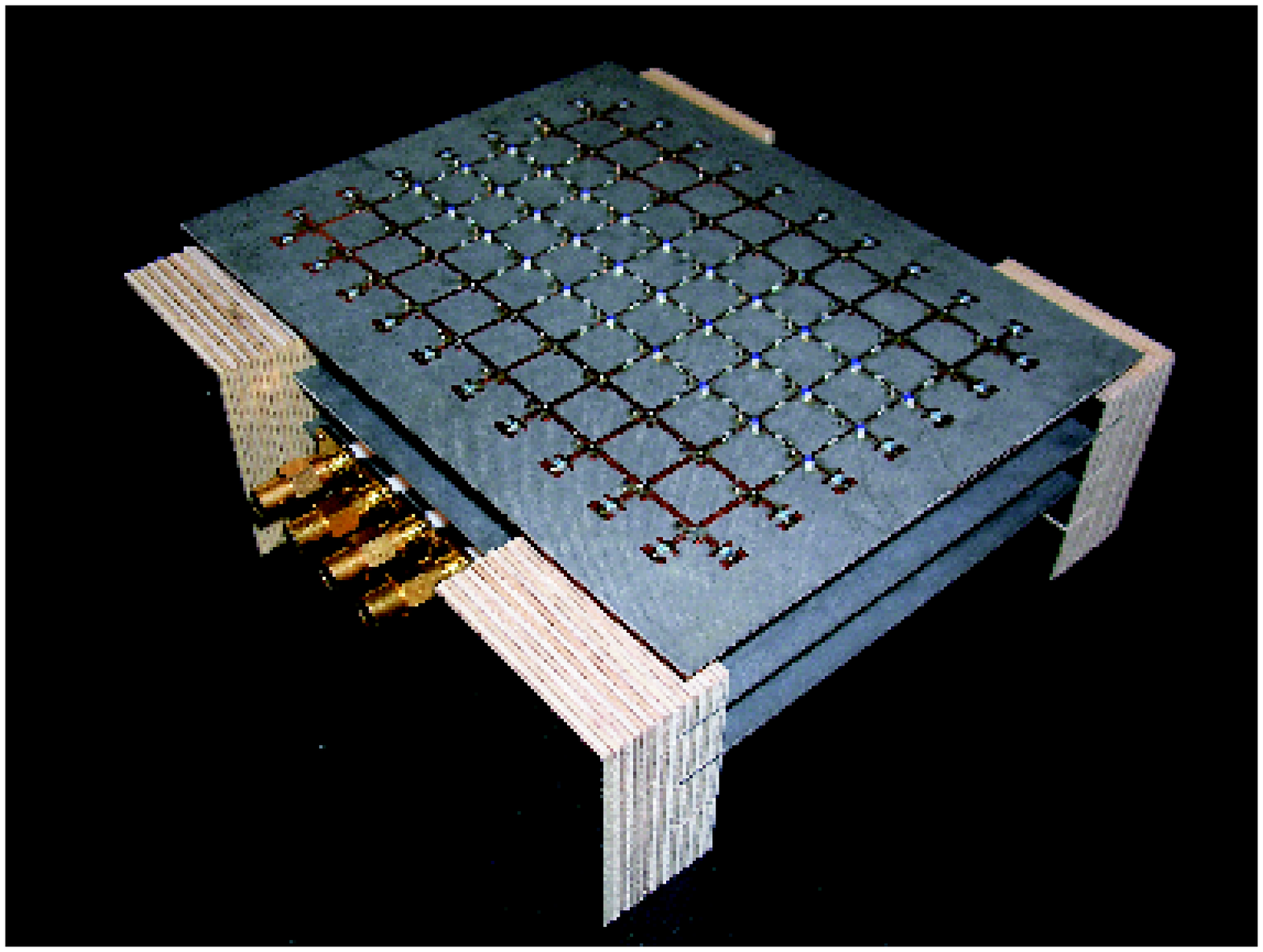, width=0.5\textwidth}
\caption{(Color online). Experimental prototype of the 3D
structure with three horizontal layers.}\label{3_layer}
\end{figure}

\begin{figure}[h]
\centering \epsfig{file=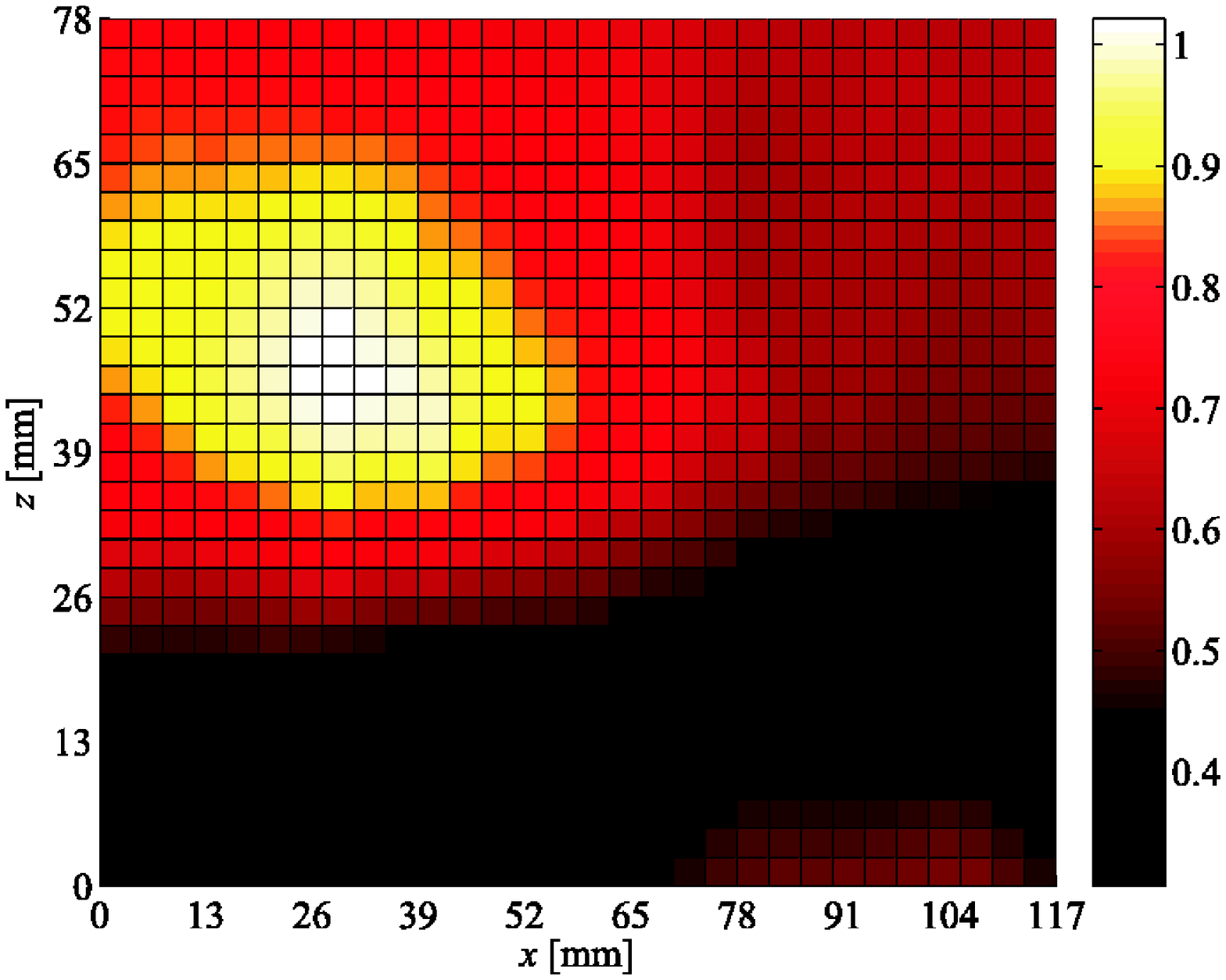, width=0.5\textwidth}
\caption{(Color online). Measured amplitude of the vertical
component of electric field on top of the 3D structure (three
horizontal layers) at $f=900$ MHz. Fields normalized to the
maximum value. One source at $x=39$ mm, $z=-6.5$ mm, situated in
the bottom layer.}\label{amplitude_3_1}
\end{figure}

Next, a third layer as in section~\ref{sec_2D_prototype} was
manufactured and appended to the top of the other two horizontal
layers using vertical sub-layers as shown in Fig.~\ref{2_layer}a.
A photograph of the manufactured three-layer structure is shown in
Fig.~\ref{3_layer}. The structure was again excited using the same
connectors as in section~\ref{sec_2D_prototype} (situated now in
the lowest horizontal layer). The electric field distribution on
top of the upper layer was measured as in
section~\ref{sec_2D_prototype}. See Fig.~\ref{amplitude_3_1} for
the measured electric field distribution on the top of the
structure.

As is seen in Fig.~\ref{amplitude_3_1}, the maximum field value
occurs near the back edge of the BW region. Vertical propagation
in the BW region was verified as in the case of two horizontal
layers and similar results as in Fig.~\ref{amplitude_2_b} and
Fig.~\ref{contours_2_b} were obtained.

\section{Conclusions}

In this paper we have described realization and testing of a
three-dimensional transmission-line network which is a circuit
analogy of the superlens proposed by Pendry. The backward-wave
slab, which is the key part of this superlens, is implemented by
loading a transmission-line network with lumped inductive and
capacitive components. Detailed theoretical analysis of such
structures will be published elsewhere.\cite{Alitalo} In this
paper we have shown that a realizable three-dimensional superlens
can be quite easily designed and manufactured. A prototype of the
designed structure has been built and backward-wave propagation
and amplification of evanescent waves in the structure have been
verified by measurements of electric field distributions.

\subsection*{Acknowledgments}
This work has been done within the frame of the
\textit{Metamorphose} Network of Excellence and partially funded
by the Academy of Finland and TEKES through the
Center-of-Excellence program. The authors would like to thank Dr.
Mikhail Lapine for helpful discussions.

\end{document}